\documentstyle[fleqn,epsfig]{revp} 
\newfont{\sff}{cmssi12} 
\newfont{\bigsf}{cmss12 scaled 2000} 
\newfont{\midsf}{cmss12 scaled 1000} 
\newfont{\smlsf}{cmss12 scaled 600}  
\newfont{\bigsff}{cmssi12 scaled 2000} 
\newfont{\sfi}{cmssi10} 

\newcommand{\pmonth}{1999}
\newcommand{\spec}{nuclear physics: NUCOLEX99 20.03.-24.03.1999}
\newcommand{\no}{}

\newcommand{\be}{\begin{eqnarray}}
\newcommand{\ee}{\end{eqnarray}}

\begin{document}
\parindent 0pt
\parskip 12pt
\setcounter{page}{1}

\title{Damping of IVGDR - Fermi-liquid or Fermi-gas ?}

\author{Klaus Morawetz,$^{*1}$ Uwe Fuhrmann,$^{*1}$ \\  
\sff
*1 Fachbereich Physik, University Rostock, D-18055 Rostock,
Germany\\}

\abst{Collisional relaxation rates of collective modes in nuclei 
      are calculated using the Levinson equation for the reduced density 
      matrix with a memory dependent collision term. 
      Linearizing the collision integral two contribution have to be 
      distinguished, the one from the quasiparticle energy and
      the one from occupation factors. The first one yields the known
      Landau formula of zero sound damping and the second one leads 
      to the Fermi gas model of Ref.$^{1)}$ with the additional factor 3
      in front of the frequencies. Adding both contribution we obtain a final
      relaxation rate for the Fermi liquid model.
      Calculations of the temperature dependence of the damping rates and 
      of the shape evolution of IVGDR  are in good agreement 
      with the experiment and show only minor differences between both models.}

\maketitle
\thispagestyle{headings}

Recently the experimental observation of temperature dependence
of IVGDR has raised much interest in theoretical investigations.
The hope is that the temperature dependence gives insight into
the character of bulk nuclear matter. In analogy to zero sound
damping the nuclear matter is thought to be described as a Fermi
liquid. However there is not paid much attention to the fact that a
Fermi liquid leads to different predictions as the Fermi gas. In
Ref.$^{2)}$ the calculation was performed within a Fermi gas
model and later corrected in an errata. The latter one was claimed to be
performed as a Fermi liquid with an additional contribution from
the quasiparticles to the Fermi gas. We will show that the quasiparticle part 
alone leads to the Landau damping of zero sound$^{3)4)}$. 

With the experimental data at hand we have the possibility to
check which behavior describes the temperature dependence more
appropriate.

In this letter we shortly sketch the derivation of the damping of
a collective mode within a Fermi gas and a Fermi liquid model. 
We will show that we get the latter one from the Fermi gas model
with an additional contribution from the quasiparticles.

We start with the Fermi gas where the dispersion relation between
momentum and energy is given by $\epsilon=p^2/2m$ and will show
later what has to be changed for a Fermi liquid where $\epsilon$
is a solution of the quasiparticle dispersion relation. We will
see that the contributions from the quasiparticles alone leads to the Landau
formula of zero sound damping$^{3)4)}$
\be
\gamma\propto \left[1+\left({\Omega\over  2\pi T} \right)^2\right]
\label{lan}.
\ee

Our considerations will start conveniently from the Levinson equation
for the reduced density matrix $f$ which is valid at short time processes
compared to inverse Fermi energy $\hbar/\epsilon_f$:
\be
\lefteqn{\partial_t f_1(t)={2 g\over \hbar^2}\int\limits_0^{\infty} 
d \tau \int {dp_2 dp_3
dp_4\over (2 \pi \hbar)^6}|T|^2} \nonumber\\
\lefteqn{\times\cos{\left(\int\limits_t^{t-\tau}\Delta\epsilon(\tau)
d \tau/\hbar\right)}\delta(\Delta p)
\left({\bar f_1}{\bar f_2}f_3f_4-f_1f_2{\bar f_3}{\bar f_4}\right)_{t-\tau} }\nonumber\\
\label{levin}
\ee
where ${\bar f}=1-f$, $\Delta p=p_1+p_2-p_3-p_4$ etc., g is the
spin-isospin degeneracy and the transition probability is given by
the scattering $T$-matrix. In case that the quasiparticle 
energies $\epsilon(t)$ become time independent like in the Fermi gas model, 
the integral in the $cos$ function reduces to the familiar 
expression $\Delta \epsilon \tau$. We linearize this collision integral
with respect to an external disturbance according to
\be
f=n+\delta f \label{lineari}
\ee
where n is the equilibrium distribution. 
Clearly two contributions have to be distinguished, 
the one from the quasiparticle energy and the one from 
occupation factors$^{1)}$. 
First we concentrate on the Fermi gas model where we have only 
the contribution of the occupation factors and will later add the 
contribution of the quasiparticle energies for Fermi liquid model. 
We obtain after Fourier transform of the time
\be
-i\Omega \delta f_1&=&\Big\langle \frac{\hbar}{2} \left[\delta_+(\Delta
\epsilon+\Omega)+\delta_-(\Delta\epsilon-\Omega)\right]
\nonumber\\
&\times&\left(\delta
F_1+\delta F_2-\delta F_3-\delta F_4\right)(\Omega)\Big\rangle.
\label{abb1}
\ee
Here we use the abbreviation
\be
\langle ...\rangle&=&\lefteqn{{2 g\over \hbar^2} \int {dp_2 dp_3
dp_4\over (2 \pi \hbar)^6}|T|^2 \delta(\Delta p)...}\nonumber\\
&=&\lefteqn{{m^3 g\over \hbar^2 (2 \pi \hbar)^6}\int d\epsilon_2
d\epsilon_3d \epsilon_4 \int {d \phi \sin \theta d\theta d \phi_2
\over \cos (\theta/2)} |T|^2...}\nonumber\\&&
\label{int}
\ee
where the last line appears from standard integration techniques
at low temperatures.
Further abbreviations are
\be
\delta_{\pm}(x)&=&\pi \delta(x)\pm i{{\cal P}\over x}\approx \pi
\delta(x)\nonumber \\
\delta F_1&=&-\delta f_1({\bar n_2}n_3 n_4 +n_2 {\bar n_3}{\bar
n_4}).
\ee
The approximation used in the first line consists in the neglect
of the off-shell contribution from memory effects. This is
consistent with the used integration technique (\ref{int}). This
terms would lead to divergences which has to be cut off$^{5)}$.

Neglecting the backscattering terms $\delta F_{2/3/4}$ we obtain
from (\ref{abb1}) a relaxation time approximations with the
relaxation time
\be
\lefteqn{{1\over \tau(\epsilon_1)}={3\over 4 \pi^2 \tau_0}
\int\limits_{-\lambda}^{\infty} d x_2 dx_3 dx_4 \left[\delta (\Delta
x+\omega)+\delta(\Delta x-\omega)\right]}
\nonumber\\
\lefteqn{\qquad\quad\times({\bar n_2} n_3 n_4+n_2 {\bar n_3}{\bar n_4})}
\label{int1}
\ee
with $\omega =\Omega/T$, $x=(\epsilon-\mu)/T$, $\lambda =\mu/T$
and the time
\be
{1\over \tau_0}={2 g m T^2 \over 3 \hbar^3} \sigma.
\ee
Here we have used the definition of cross
section $|T|^2=(4 \pi\hbar^2/m)^2d\sigma/d\Omega$ and have
assumed a constant cross section $\sigma$.
To calculate (\ref{int1}) one needs the standard integrals for
large ratios of chemical potentials $\mu$ to temperature $\lambda=\mu/T$
\be
&&\int\limits_{-\lambda}^{\infty} d x_2 dx_3 dx_4 n_2 {\bar
n_3}{\bar n_4} \delta(\Delta x \pm \omega)
\nonumber\\
&&=\frac 1 2 {\bar
n_1}(x_1\pm\omega)\left[\pi^2+(x_1\pm\omega)^2\right]
\label{stand}
\ee
to obtain
\be
{1\over \tau(\epsilon_1)}= {3 \over 8 \pi^2 \tau_0} \left[2
\pi^2+(x_1+\omega)^2+(x_1-\omega)^2\right].
\label{t1}
\ee
Further we employ a thermal averaging in order to obtain the mean
relaxation time finally\footnote{One has to use the identities
valid up to $o(\exp[-\lambda])$
\[
\int\limits_{-\lambda}^{\infty} d x n {\bar n}=1; \qquad
\int\limits_{-\lambda}^{\infty} d xx^2 n {\bar n}={\pi^2\over 3}.
\]}
\be
{1\over \tau_{\rm gas}}=\int\limits_{-\lambda}^{\infty} d x_1 n_1 {\bar
n_1} {1\over \tau(\epsilon_1)}={1\over \tau_0}\left[1+3\left({\omega\over 2
\pi}\right)^2\right].
\label{avt}
\ee
If we do not use the thermal averaging but take 
(\ref{t1}) at the Fermi energy $\epsilon_1=\epsilon_f$ we will obtain
\be
{1\over \tau(\epsilon_f)}={1\over \tau_0}\left[\frac 3 4 +3\left(
{\omega\over 2\pi}\right)^2\right].
\label{tep}
\ee
We see that both results disagree with the Landau result of
quasiparticle damping (\ref{lan}) by factors of 3 at different
places$^{1)2)8)9)}$. 
We have point out that the result at fixed Fermi energy will
lead to unphysical results for the Fermi liquid case. Therefore
we consider the thermal averaged result as the physical one.

We now turn to the Fermi liquid model and replace the free
dispersion $\epsilon=p^2/2m$ by the quasiparticle energy
$\epsilon_p$. Than the variation of the collision integral gives
an additional term which comes from the time dependence of the
quasiparticle energy on the $\cos$-term of (\ref{levin}). We have
instead of the sum of two complex conjugate exponentials in
(\ref{levin}) an additional contribution from the linearization
of the exponential
\be
&&\delta \exp{\left (i\Delta\int\limits_t^{t-\tau} d {\bar t }
\epsilon({\bar t}) \right )}\nonumber\\
&&={\rm e}^{-i\Delta \epsilon \tau}
\left
(1-i \Delta
\int\limits_t^{t-\tau} d{\bar t}[\epsilon({\bar
t})-\epsilon]
\right )
-{\rm e}^{-i\Delta \epsilon \tau}
\nonumber\\
&&={i T\over \hbar} {\rm e}^{-i\Delta \epsilon \tau/\hbar}
\Delta\int\limits_t^{t-\tau}
d{\bar t } {\delta f({\bar t})\over n {\bar n}}.
\ee
In the last line we have replaced the variation in the
quasiparticle energy $\epsilon(t)-\epsilon$
by the variation in the distribution
function $\delta f$ due to the identity$^{4)}$
\be
\delta
f(t)&=&f(t)-n(\epsilon)=f(t)-n(\epsilon(t))-[n(\epsilon)-n(\epsilon(t))]
\nonumber\\
&\approx &-n'(\epsilon(t)-\epsilon)={n{\bar n}\over
T}\left[\epsilon(t)-\epsilon\right]
\ee
where we assumed within the quasiparticle picture that
$f(t)=n(\epsilon(t))$. This leads now to an additional part in
the relaxation time which we write analogously to (\ref{abb1})
\be
{1\over \tau_c(\epsilon_1)}&=&\left\langle \frac{\hbar}{2} {\delta_+(\Delta
\epsilon-\Omega)-\delta_+(\Delta\epsilon+\Omega)\over
\Omega}\right.\nonumber\\
&&\left.\times{{\bar n_1}{\bar n_2}n_3 n_4-n_1 n_2 {\bar n_3}{\bar
n_4}\over n_1 {\bar n_1}} \right\rangle.
\label{abb2}
\ee
Using again (\ref{stand}) we obtain
\be
\lefteqn{
{1\over \tau_c(\epsilon_1)}={-3\over 4 \pi^2\tau_0}
\left \{ \frac{{\bar n}(x_1+\omega)\left[\pi^2+(x_1+\omega)^2\right]}
{2\omega({\rm e}^{-\omega}-1)}\right.}\nonumber\\
\lefteqn{\qquad\quad+\,[\omega \leftrightarrow - \omega] \Bigg\} } 
\label{tauepsq}
\ee
and get after thermal averaging (\ref{avt})\footnote{Here one uses
[$o(\exp[-\lambda])$]
\[
\int\limits_{-\lambda}^{\infty} dx n=\lambda;\quad
\int\limits_{-\lambda}^{\infty} dx x n=-\frac 1 2 \lambda^2+\frac
1 6 \pi^2;\quad
\int\limits_{-\lambda}^{\infty} dx x^2 n=\frac 1 3 \lambda^3
\].}
\be
{1\over\tau_c}={1\over \tau_0}\left[1+
\left({\omega\over 2 \pi}\right)^2\right].
\label{avt1}
\ee
Taking instead of thermal averaging the value at Fermi
energy $(\epsilon_1=\epsilon_f)$ in (\ref{tauepsq}) we find
\be
{1\over \tau_c(\epsilon_f)}={3 (\pi^2+\omega^2)\over 2
\pi^2\tau_0\omega} {{\rm e}^{\omega}-1\over {\rm e}^{\omega}+1}.
\label{tep1}
\ee
Here we like to point out that the Landau result (\ref{lan}) 
appears in (\ref{avt1}) (see also in Ref.$^{2)6)-9)}$).

Adding now (\ref{avt}) and (\ref{avt1}) we obtain a final relaxation time for
the Fermi liquid model
\be
{1\over \tau_{\rm liq}}={2\over \tau_0} \left[1+2 \left({\omega\over 2
\pi}\right)^2\right].\label{avtliq}
\ee
which is the main result in this paper.
It contains the typical Landau result of zero sound (\ref{lan}) except 
the factor 2 in front of the frequencies.
Comparing (\ref{avtliq}) with the Fermi gas model (\ref{avt}) we
see that in the limit of vanishing temperature the Fermi liquid value is
lower with $\propto 2 \Omega^2$ compared to the Fermi gas $\propto 3
\Omega^2$. 
Further for vanishing frequencies (neglect of memory effects)
the Fermi liquid model leads to twice the relaxation rate than the Fermi gas
model.  The coefficient of temperature increase is than twice
larger for the Fermi liquid than for the Fermi gas. 
If we consider the relaxation times at Fermi energy (no thermal averaging)
(\ref{tep}) and (\ref{tep1}) we find the same results as above 
in the limit of vanishing frequencies. For vanishing temperature 
only the Fermi gas (\ref{tep})  coincides  with the result of (\ref{avt})
$\propto 3\Omega^2$. Expression (\ref{tep1}) goes to zero for $T=0$
and underlines the necessity to thermal average the value.  

Next we like to apply the results (\ref{avt}) and (\ref{avtliq}) for
the calculation of the temperature dependence of damping rates of
isovector giant dipole resonances (IVGDR). This was done in Ref.$^{1)}$
where the main result (Eq.(52)) corresponds the   Fermi gas result
(\ref{avt}). 
The linearization of the kinetic equation (\ref{levin}) 
according to (\ref{lineari}) leads to an extended polarization function of
Mermin$^{10)}$ 
\be
\Pi^{\rm M}_0({\bf q},\omega)=
\frac{\Pi_0({\bf q},\omega+{i/\tau})} {
1-\displaystyle{\frac{i}{\omega\tau +i}} \left[1- 
\displaystyle{\frac{\Pi_0({\bf q},\omega+{i/\tau})}{ \Pi_0({\bf q},0)}} \right]}
                \label{merminDF},
\ee 
in order to incorporate the collision effects (\ref{avt}) and (\ref{avtliq}) 
into the polarization function $\Pi_0$.
The energy and damping rates are now determined by the zeros of the
(Mermin) polarization function. 
\be
\epsilon^M({\bf q},\Omega+i \gamma)=
1-V\Pi_0^{\rm M}({\bf q},\Omega+i \gamma)=0.
\label{eps_Merm}
\ee
With (\ref{eps_Merm}) we have for the strength function
\be
S({\bf q},\omega)=\displaystyle{\frac 1 \pi {{\rm Im}\Pi_0
                  \over(1-V {\rm Re}\Pi_0)^2+(V{\rm Im}\Pi_0)^2}}.
       \label{strenght}
\ee
  
\begin{figure}[h]
\centerline{\epsfig{file=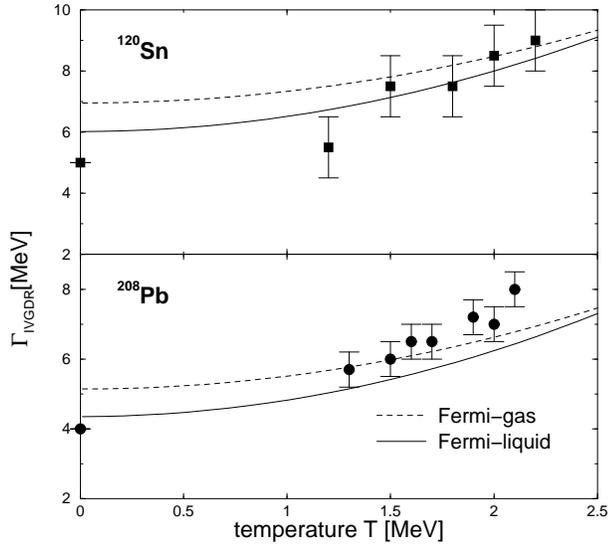,scale=.4,angle=-90}}
\vspace{.2cm}
 \caption{Experimental damping rates  vs. temperature of IVGDR 
          for $^{120}$Sn and $^{208}$Pb 
          ($^{120}$Sn from Ref.$^{11)}$ and $^{208}$Pb from
          Ref.$^{12)}$) compared with the solution of the dispersion
          (\ref{eps_Merm})
          relation $\Gamma=2\,\gamma$  for the Fermi gas (dashed lines) and
          the Fermi liquid model (solid lines).}
                      \label{bild1}
\end{figure}
In Fig. \ref{bild1} we have plotted the theoretical 
damping rates $\Gamma=2\,\gamma$
of the IVGDR modes in $^{120}$Sn and 
$^{208}$Pb as a function of temperature together with experimental values.
As long as the results of the Fermi gas model (\ref{avt}) and 
the Fermi liquid model (\ref{avtliq}) are very close and in good
agreement with the data the temperature dependence still remains too flat 
compared to the experiments. The small difference between both models for T=0 
vanishes with increasing temperature.  
\begin{figure}
\centerline{\epsfig{file=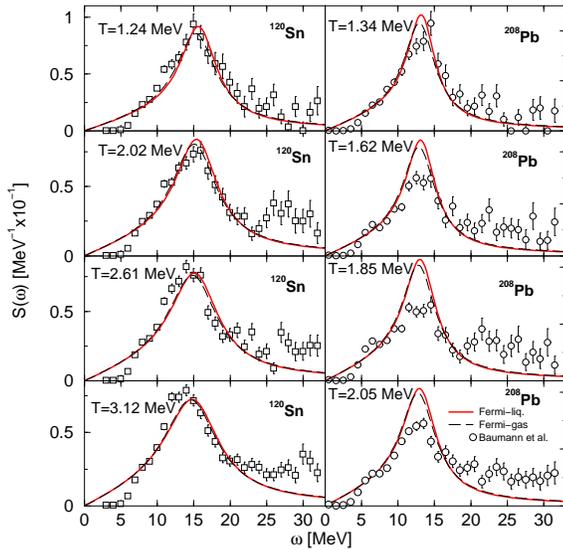,scale=.45}}
\vspace{.2cm}
 \caption{The IVGDR strength function in  $^{120}$Sn (LHS) and $^{208}$Pb (RHS)
          within the Fermi gas model (dashed lines) and Fermi liquid model 
         (solid lines) at serveral temperatures compared with normalized
          data from Ref.$^{13)}$. }
                      \label{bild2}
\end{figure}
Comparing also the observerd shape evolution of IVGDR strength function with
our models underlies the latter fact.
In Fig. \ref{bild2} we have plotted the strength function (\ref{strenght})
for $^{120}$Sn (LHS) and $^{208}$Pb (RHS)
within the Fermi gas model (\ref{avt}) (dashed lines) and 
Fermi liquid model (\ref{avtliq}) (solid lines) with the normalized data from
Ref.$^{13)}$.
The good overall agreement of the shape evolution of both models with the
experiment is again accompanied  with only minor differences between the
Fermi gas and the Fermi liquid model.

\section*{References}
\re
1) U.\ Fuhrmann, K.\ Morawetz and R.\ Walke: Phys.\ Rev.\ C\ {\bf 58}, 
1472 (1998).
\re
2) S.\ Ayik and D.\ Boilley: Phys.\ Lett.\ B\ {\bf 276},  263  (1992): 
ibid.\ {\bf 284},  482(E) (1992).
\re
3) A.\ A.\ Abrikosov and I.\ M.\ Khalatnikov: Rep.\ Prog.\ Phys.\ {\bf 22}, 
329 (1959).
\re
4) E.\ M.\ Lifshitz and L.\ Pitajevski: 
{\em Physical Kinetics} (Nauka, Moscow, 1978).
\re
5) K.\ Morawetz and H.\ S.\ K\"ohler: Europhys.\ J.\ A (1999), in press.
\re
6) S.\ Ayik, M.\ Belkacem, and A.\ Bonasera: Phys.\ Rev.\ C\ {\bf 51}, 
611 (1995).
\re
7) S.\ Ayik, O.\ Yilmaz, A.\ Gokalp, and P.\ Schuck: Phys.\ Rev.\ C\ {\bf 58}, 
1594  (1998).
\re
8) V.\ M.\ Kolomietz, A.\ G.\ Magner, and V.\ A.\ Plujko:
Z.\ Phys.\ A\ {\bf 345}, 131 (1993).
\re
9) A.\ G.\ Magner, V.\ M.\ Kolomietz, H.\ Hofmann, and S.\ Shlomo: 
Phys.\ Rev.\ C\ {\bf 51}, 2457 (1995).
\re
10) N.\ D.\ Mermin: Phys.\ Rev.\ B\ {\bf 1},  2362  (1970).
\re
11) E.\ Ramakrishnan et al.: Phys.\ Rev.\ Lett.\ {\bf 76},  2025  (1996).
\re
12) E.\ Ramakrishnan et al.: Nucl.\ Phys.\ {\bf A549},  49  (1996).
\re
13) T.\ Baumann et al.: Nucl.\ Phys.\ {\bf A635},  428  (1998).

%
%
%
%
%
%

\end{document}